\documentclass[review]{elsarticle}

\usepackage{lineno,hyperref}
\modulolinenumbers[5]

\journal{Astroparticle Physics}

\usepackage{graphicx}
\usepackage{subfigure}
\usepackage{multirow}
\usepackage{soul,color}
\setstcolor{red}

\begin{document}
\graphicspath{{Figure/}}

\begin{frontmatter}

\title{Slow Liquid Scintillator Candidates for MeV-scale Neutrino Experiments}

\author[THU,LAB]{Ziyi~Guo\corref{cor1}}
\author[BNL]{Minfang~Yeh}
\author[NJU]{Rui~Zhang}
\author[NJU]{De-Wen~Cao}
\author[NJU]{Ming~Qi}
\author[THU,LAB]{Zhe~Wang\corref{cor2}}
\author[THU,LAB]{Shaomin~Chen}

\cortext[cor1]{Corresponding author: guo-zy17@mails.tsinghua.edu.cn}
\cortext[cor2]{Corresponding author: wangzhe-hep@mail.tsinghua.edu.cn}

\address[THU]{Department~of~Engineering~Physics, Tsinghua~University, Beijing 100084, China}
\address[LAB]{Key~Laboratory~of~Particle~\&~Radiation~Imaging (Tsinghua~University), Ministry~of~Education, China}
\address[BNL]{Brookhaven~National~Laboratory, Upton, New York 11973, USA}
\address[NJU]{School of Physics, Nanjing University, Nanjing 210093, China}

\begin{abstract}
Slow liquid scintillator Cherenkov detectors have been proposed as part of several future neutrino experiments because they can provide both directionality and energy measurements.
This feature is expected to enhance the sensitivities for MeV-scale neutrino physics, including solar physics, the search for supernova relic neutrino, and the study of geo-sciences.
In this study, the characteristics of a slow liquid scintillator were investigated, along with the light yields and decay time constants for various combinations of linear alkylbenzene (LAB), 2,5-diphenyloxazole (PPO), and 1,4-bis (2-methylstyryl)-benzene (bis-MSB).
The results of our study indicated that LAB with 0.07\,g/L of PPO and 13\,mg/L of bis-MSB was the best candidate for an effective separation between Cherenkov and scintillation lights with a reasonably high light yield.
\end{abstract}

\begin{keyword}
Cherenkov \sep scintillation \sep slow liquid scintillator \sep neutrino detection
\end{keyword}

\end{frontmatter}

\newcommand{\ud}{\mathrm{d}}
\newcommand{\ui}{\mathrm{i}}
\newcommand{\ue}{\mathrm{e}}
\section{Introduction\label{sec:intro}}
The China Jinping Underground Laboratory (CJPL) is located in Sichuan Province, China. With an overburden of about 2400\,m~\cite{kang2010status} and located approximately 1000\,km from the closest nuclear power plant, the CJPL is an ideal site for low background and MeV-scale neutrino experiments.
The Jinping Neutrino Experiment was proposed at CJPL~\cite{beacom2017physics,geo-WanLinyan, geo-Bill} with the primary goals of focusing on solar neutrinos, geoneutrinos, and supernova relic neutrinos (also referred to as the diffuse supernova neutrino background).

In MeV-scale low-energy neutrino experiments, the directional information of charged particles can be reconstructed by Cherenkov light. A number of studies~\cite{beacom2017physics, alonso2014advanced, YEH201151} have indicated that directional information (for particle indentification) and an accurate measurement of the energy on charged particles may provide extra discriminating power to the background suppression in MeV-scale neutrino experiments.
For example, the solar angle cut on the direction of charged particles is a powerful selection criterion for solar neutrino events.
Another study~\cite{wei2017discovery} demonstrated that atmospheric neutrino background via the neutral and charged current interactions, which is one of the major backgrounds in the search of supernova relic neutrino events, can be effectively suppressed if electrons and muons are distinguished from non-Cherenkov produced neutrons and protons by particle identification.

In addition, it is possible to further perform particle identification based on the ratio of Cherenkov light yield to scintillation light yield. Both features are also useful for neutrinoless double beta decay experiments~\cite{JINST,Fukuda:2016yjg,elagin2017separating}, neutrino CP phase measurements~\cite{ciuffoli2016neutrino, ciuffoli2014leptonic}, proton decay searches~\cite{YEH201151} and the study of geoneutrinos~\cite{geoZhe}.

The detection scheme with slow scintillator is now under consideration in the Jinping Neutrino Experiment~\cite{beacom2017physics} and THEIA~\cite{alonso2014advanced}.
Although the concept of diluted scintillators was pioneered as part of the LSND experiment~\cite{LSND}, its low light yield is not applicable to dedicated low-energy neutrino experiments. It is import to note that the discriminating individual solar neutrino flux component from the recoiled electron energy spectrum requires a reasonable energy resolution, which should be at least $4.5\%$ at a 1\,MeV energy deposit, i.e., 500 photoelectrons/MeV)~\cite{beacom2017physics}.
This requirement exceeds the yield limit of photoelectrons in water or heavy water Cherenkov detectors.
While liquid scintillator detectors can meet this requirement, those adopted in present neutrino experiments only provide energy information, because the extremely small amount of Cherenkov light emitted by charged particles is completely submerged by the huge scintillation light.
A new type of slow liquid scintillator (water- or oil-based) has much larger time constants, which provides the opportunity to separate the Cherenkov light from the scintillation light. The concept of water-based liquid scintillators (WbLS) was proposed as early as the findings in~\cite{winn1985water} and recent efforts toward achieving this can be found in~\cite{YEH201151, bignell2015characterization, bignell2015measurement, So:2014hua}.

Linear alkylbenzene (LAB) was revisited in~\cite{li2016separation}.
A quadruple coincidence system was used to select vertical cosmic-ray muons.
It was observed that LAB has a characteristic of a large decay time constant of 35\,ns, and is therefore classified as a slow liquid scintillator.
This feature could be used to separate Cherenkov and scintillation lights by analyzing the time profile of the analogous output of a photomultiplier tube (PMT), given that the prompt time region will be dominated by the former while the later times will be dominated by the latter.
The slower the fluorescence, the better the separation ability.

Since the light yield of LAB is much lower than that of typical liquid scintillators widely used in neutrino experiments, especially in low-energy solar neutrino experiments, CHESS experiment~\cite{CHESS,CHESS2} has made a good progress by adding 2,5-diphenyloxazole (PPO) and applying fast photon detectors~\cite{LAPPD} to enhance the light yield and maintain the scintillation-Cherenkov separation ability.
However, for large neutrino detectors (quick absorption below 400 nm in LAB) or neutrino detectors using acrylic material (transmittance is cut off at about 300~nm), the light propagation loss cannot be ignored. We still need to shift the emission spectrum from the short wavelength to the longer range ($>400$\,nm) to reduce the light propagation loss. This can be achieved by adjusting the concentrations of PPO and 1,4-bis (2- methylstyryl)-benzene (bis-MSB)~\cite{DYB-LS1,DYB-LS2}.
On the other hand, adding too much PPO and bis-MSB will weaken the scintillation and Cherenkov separation ability because the Cherenkov photons will be absorbed and the time constants will be smaller. This is especially critical when using a more economical PMT detection approach (the timing precision of PMT is on the nanosecond scale and massive production is possible).
The balance of time profile and light yield is vital to both the Cherenkov separation ability and high energy resolution and thus requires further study.

In this study, we investigated the effect of adding both PPO and bis-MSB to LAB.
We first scanned the light yields and time profiles of cosmic-ray muons for various LAB with PPO and bis-MSB combinations, and used an energy transfer model to describe the inverse relationship between them (Section~\ref{sec:Yield-Time}). We then measured the scintillation emission spectra (Section~\ref{sec:emission}) and the transmissions in acrylic (Section~\ref{sec:acrylic}),which is a typical material for scintillator containers.
We further measured the attenuation length (Section~\ref{sec:attenuation}) for a typical sample with a long-arm apparatus. We evaluate the performances of candidate samples for neutrino detection in Section~\ref{sec:Candidate}, and finally, we summarize the findings of our study in Section~\ref{sec:conclusion}.
\section{Study of light yield and scintillation time\label{sec:Yield-Time}}
\subsection{Apparatus\label{App}}
\begin{figure}
\centering
\includegraphics[width=0.7\textwidth]{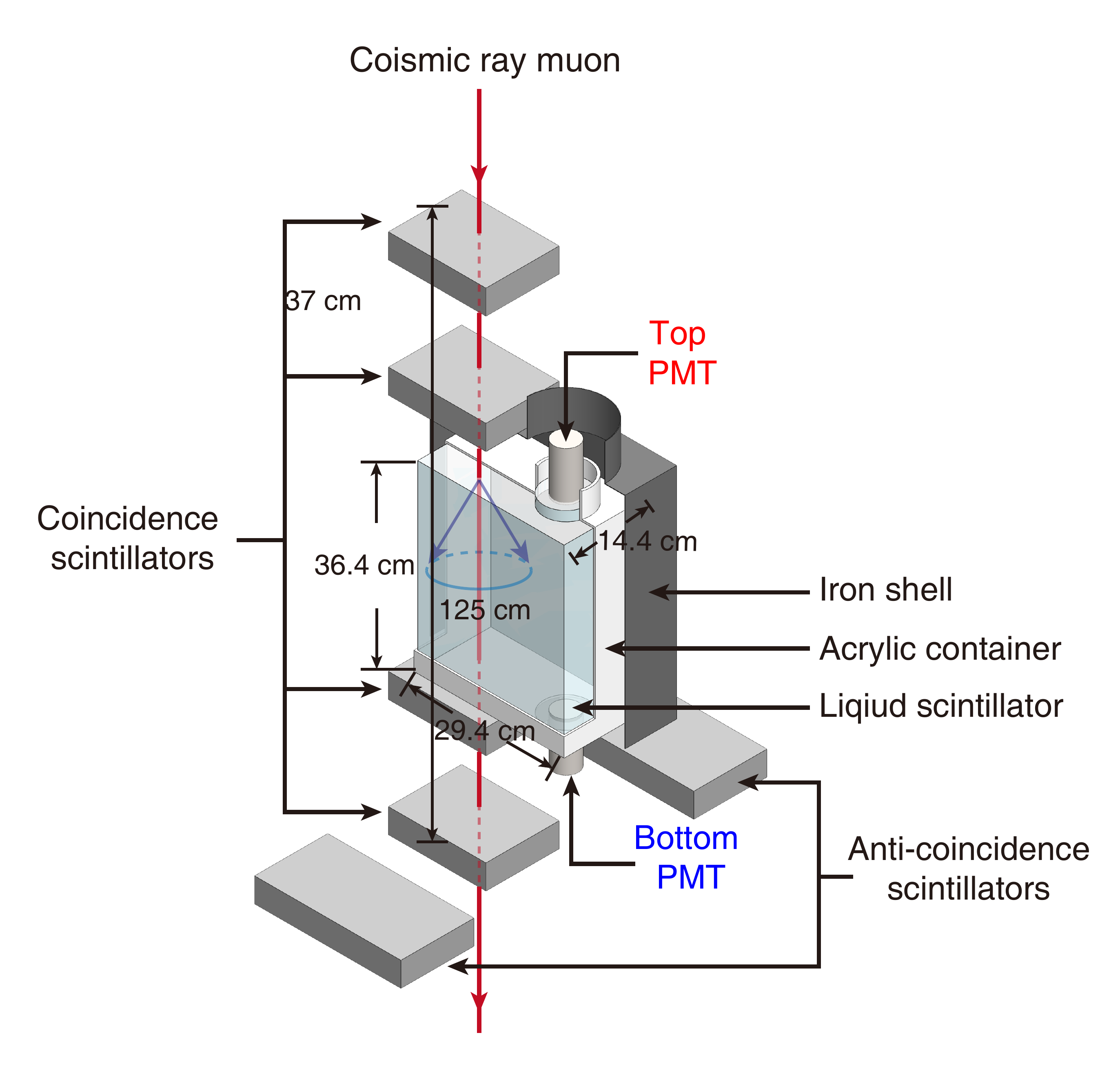}
\caption{Detector structure for this slow liquid scintillator study.\label{front}}
\end{figure}

The detector is shown in Fig.~\ref{front}; in the setup, where four plastic scintillators were positioned vertically. The four coincident signals were used to provide a trigger for a muon traveling from the top to the bottom. Two additional plastic scintillators were placed next to the coincident bottom scintillator to serve as a veto counter to exclude events with muon shower activity, which can destroy the distinguished time profile.

Approximately 15.4\,L of a liquid scintillator sample was placed in an acrylic container (36.4\,cm height) and the container was placed between the second and third coincident scintillators. PPO and bis-MSB were then weighed by an electronic balance (1\,mg division minimal) and dissolved in a 500\,mL beaker filled with LAB. The  concentrated solution was poured into the acrylic container, and the mixture was thoroughly stirred.

The inner surface of the acrylic container was lined by a layer of black coarse acrylic to suppress reflections. The top and bottom PMTs symmetrically aligned with the acrylic container were then immersed in the liquid scintillator.

The light signals from the six plastic scintillators and the liquid scintillator were collected by eight PMTs. The top and bottom PMTs used to acquire the signals of the liquid scintillator were Hamamatsu model R1828-01 with a 46\,mm diameter for the effective photocathode area.
The quantum efficiency is more than 10\% from 300 to 530\,nm.
The rise time of the anode pulse is 1.3\,ns.
Other parameters of the PMTs can be found elsewhere~\cite{PMT}.

Once a trigger signal was issued, a 10\,bit, 1\,GHz flash analog-to-digital converter (FADC, model CAEN V1751) opened a 4096-ns window and read out the voltage waveforms for all eight PMTs.

\subsection{Event selection}
To eliminate electronic noise and multi-track events, several characteristic variables, namely peak, width and charge (waveform area), were studied for each waveform.

\begin{itemize}
\item Electronic noises were primarily caused by the power supply of the PMTs. The waveform of this noise was much narrower than that produced by a photon. Cuts on both the peak-to-charge-ratio and peak-to-width-ratio were applied to significantly suppress these backgrounds.
\item Muon showers may occur when an energetic muon spalls with an atom. This background was first rejected by the two anti-coincidence scintillators. The charges of the four coincidence scintillators were also applied when an event with charge significantly deviated from the average.
\end{itemize}

\begin{figure}
\subfigure[\ The average waveform of the top PMT. Only the scintillation light is presented. \label{waveform1}]{
\includegraphics[width=0.45\textwidth]{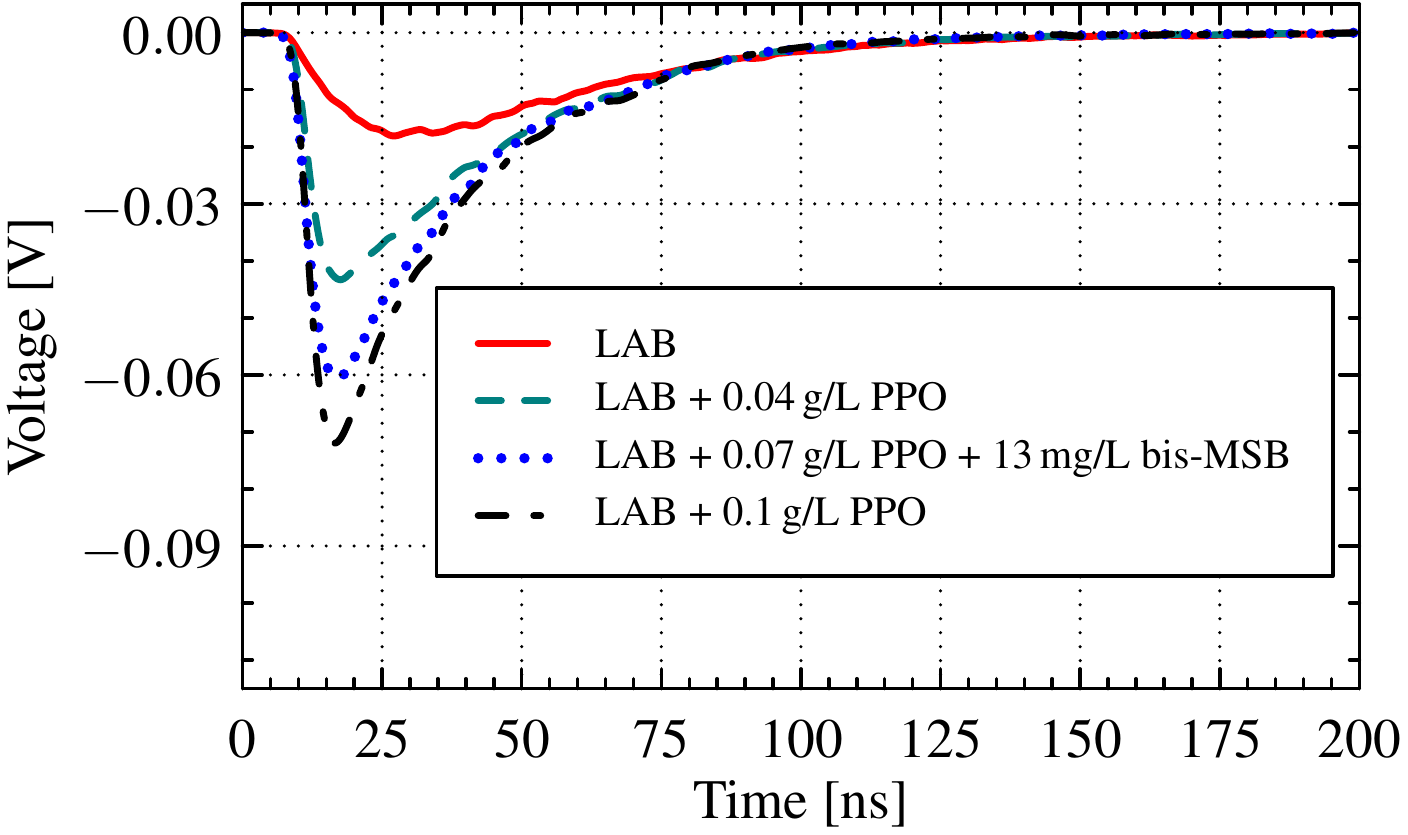}}
\subfigure[\ The average waveform of the bottom PMT. The sum of Cherenkov and scintillation lights are presented. \label{waveform2}]{
\includegraphics[width=0.45\textwidth]{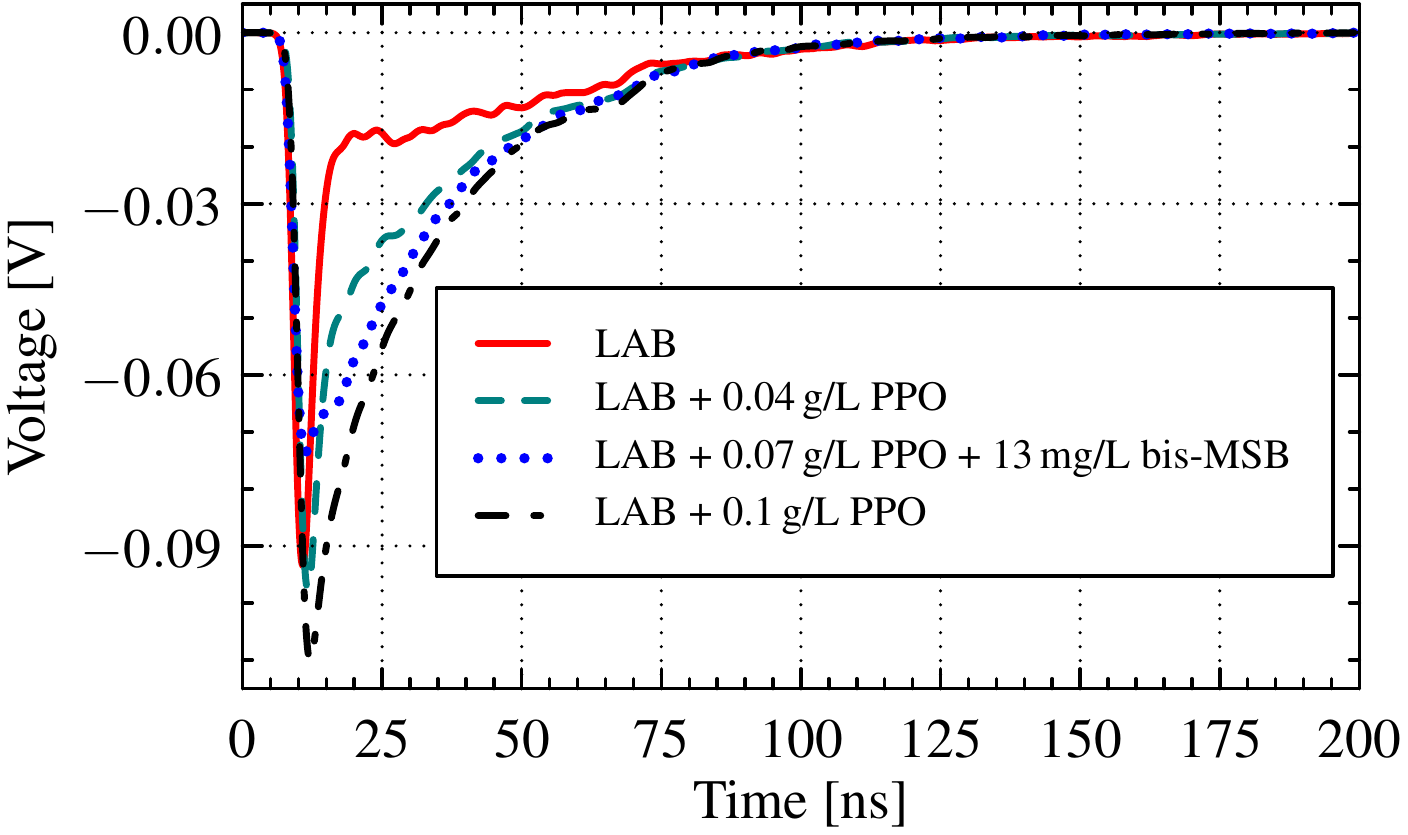}}
\caption{The average waveforms of the top and bottom PMTs.}
\end{figure}

For the purpose of comparison, we selected 2,000 candidates for the top and 2,000 for the bottom PMTs. The event rate was about 1.7\,/min. The average waveforms among the selected candidates were then calculated, as shown in Figures~\ref{waveform1} and \ref{waveform2}, respectively.
The relative difference of the gains and acceptances between the two PMTs were corrected by the gain calibration and Monte-Carlo simulation in these figures.
Since the vertical muons only come from the top, the top PMT can only detect the isotropic scintillation light rather than the forward Cherenkov light emitted by these muons, while the bottom PMT can detect both lights. We investigated the time profiles from the two PMTs and observed that there was a clear enhancement in the first 20\,ns for the amplitude of the bottom PMT with respect to the top one.
This is due to the contribution from the prompt Cherenkov component and the peak height is highly dependent on the concentration of PPO and bis-MSB.
A previous study~\cite{Johnny} indicated that the absorption of light is intensive for those with wavelength less than 400\,nm, and the number of Cherenkov photons is reduced as a consequence

\subsection{Time profile measurement}
We constructed a function of the time profile with the PMT time response convoluted,  taking both the Cherenkov and scintillation light contributions into account, as expressed by,
\begin{equation}
f_b(t)=
\left[A_c\delta(t-t_c)+A_sn(t-t_{s})\right]\otimes\textrm{gaus}(\sigma_b),
\label{waveformb}
\end{equation}
where $A_{c}$ is the amplitude of the Cherenkov light, $t_c$ is the mean arrival time of the Cherenkov light, $\delta(t)$ represents the time profile of the prompt Cherenkov emission, which is a delta function since it is an instant process comparing to PMT timing precision of ns. $A_{s}$ and $t_s$ are the amplitude and start time of the scintillation light, respectively, $n(t)$ is the time profile of the scintillation emission, $\textrm{gaus}(\sigma_b)$ is the PMT time response function.
In a binary or ternary scintillator system, emissions may feature a finite rise time or be slightly lengthened in duration due to the finite time of intermolecular energy transfer~\cite{Birks}.
In organic solution scintillators, emissions present a finite rise time $\tau_r$ and a decay time $\tau_d$ so that a normalized pulse shape of scintillation light can be written as
\begin{equation}
n(t)=\frac{\tau_r+\tau_d}{\tau_d^2}(1-\mathrm{e}^{-t/\tau_r})\cdot\mathrm{e}^{-t/\tau_d}
\label{timeprofile}
\end{equation}

\begin{figure}
\subfigure[\ The top PMT. \label{fitt}]{
\includegraphics[width=0.45\textwidth]{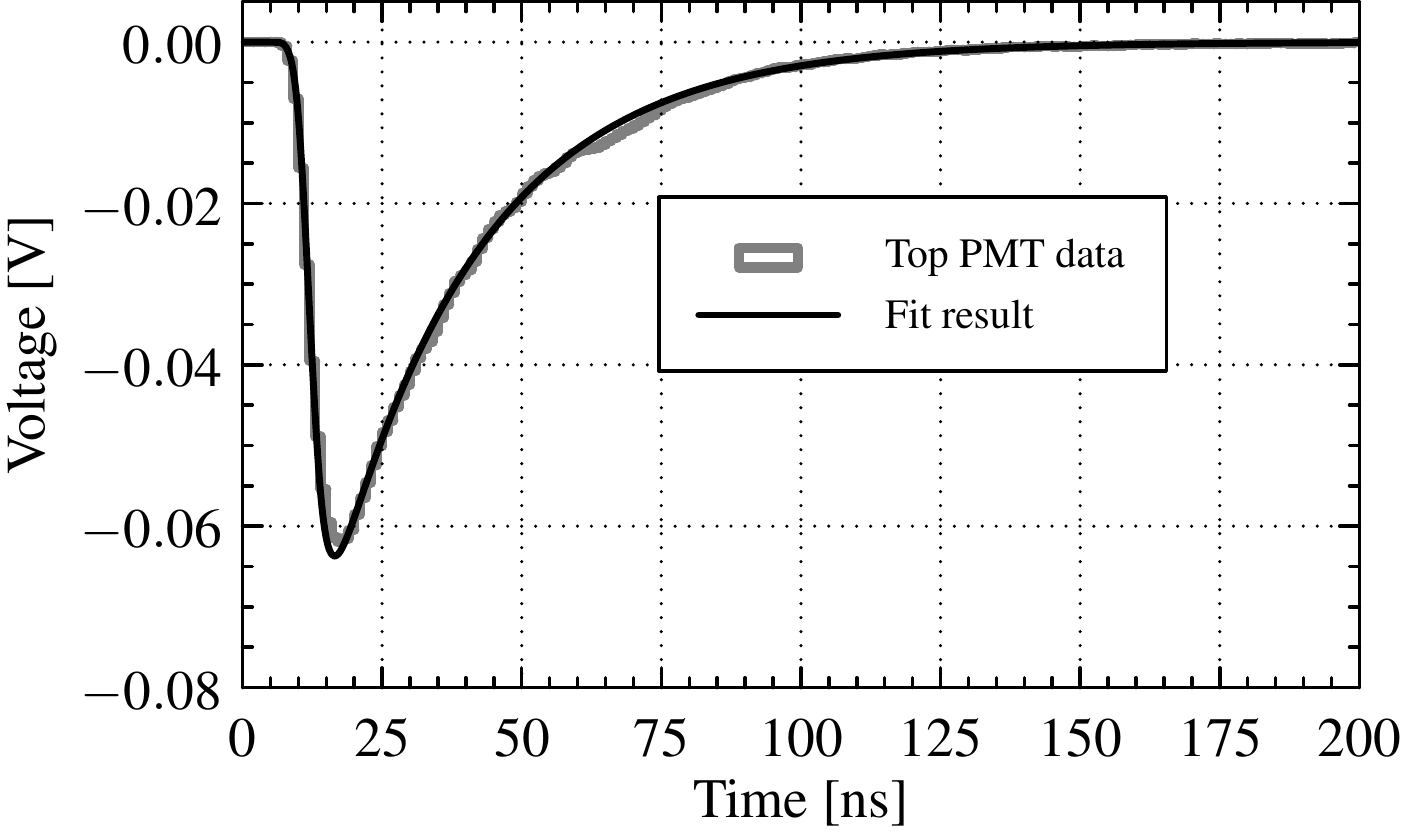}}
\subfigure[\ The bottom PMT. \label{fitb}]{
\includegraphics[width=0.45\textwidth]{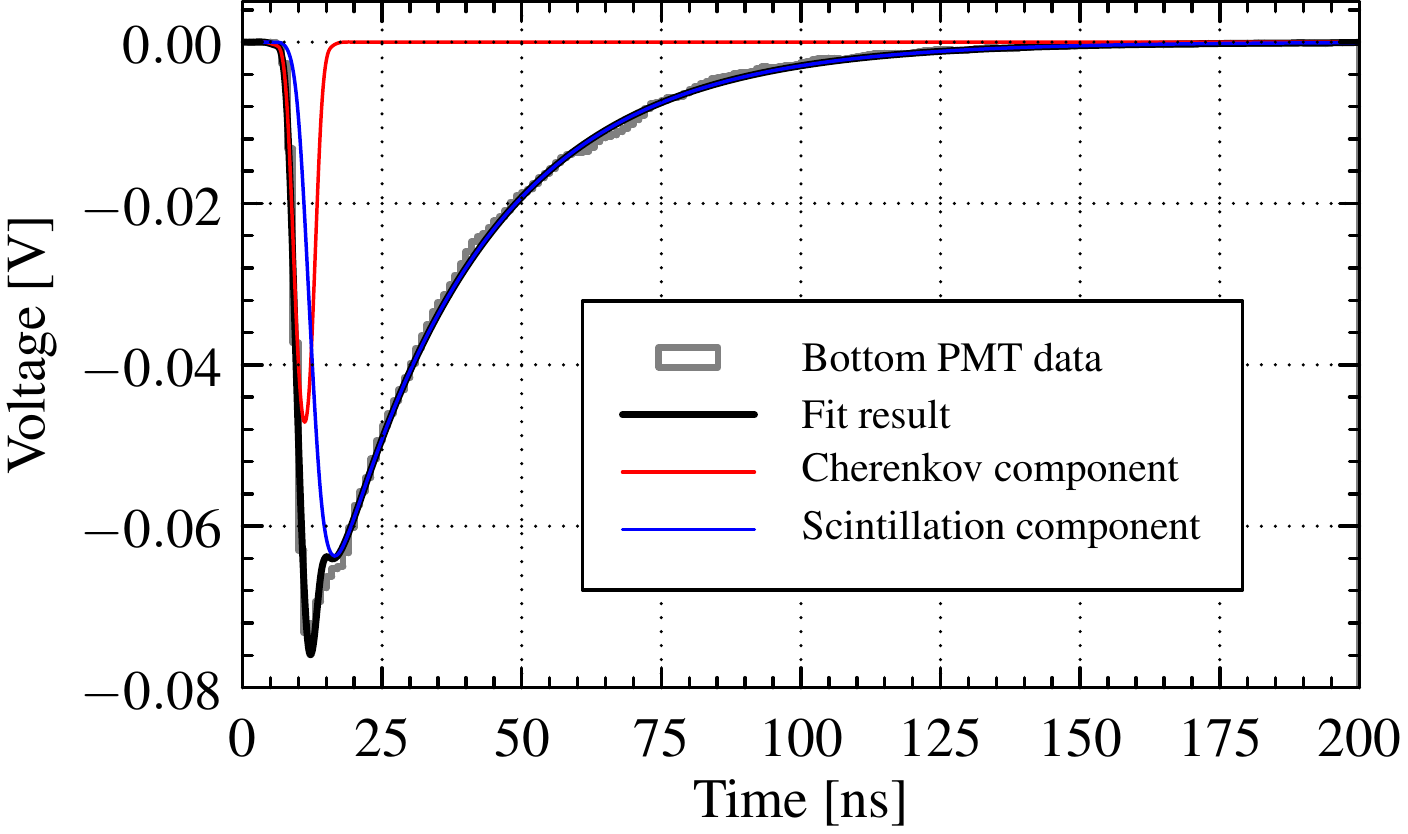}}
\caption{The fit to the waveforms of the top and bottom PMT for the sample of LAB with 0.07~g/L PPO and 13~mg/L bis-MSB. \label{fig:fit}}
\end{figure}

In contrast to the waveform of the bottom PMT, the waveform of the top PMT includes only the scintillation light contribution, which is expressed as
\begin{equation}
f_t(t)=A_s n(t-t_{s})\otimes \textrm{gaus}(\sigma_t).
\label{waveformt}
\end{equation}
where $A_s$ is the scintillation amplitude, $t_s$ is the start time of the scintillation light, which are the same as the values in Eq.~(\ref{waveformb}), and $n(t)$ is the time profile in Eq.~(\ref{timeprofile}). The time resolution $\sigma_t$ is also taken into account.
Both the time constants $\tau_r$ and $\tau_d$ can be determined by Eq.~(\ref{waveformb}) and Eq.~(\ref{waveformt}).
For example, from a LAB sample with 0.07 g/L PPO and 13 mg/L bis-MSB we determined $\tau_r=(1.16\pm0.12)$~ns and $\tau_d=(26.76\pm0.19)$~ns, respectively. The fitting results for both the top and bottom PMT waveforms are shown in Fig.~\ref{fitt} and \ref{fitb}, respectively.

\subsection{Light yield measurement}

The number of scintillation photoelectrons (PE) $D_s$ detected by the bottom PMT can be expressed as,
\begin{equation}
D_{s}=\frac{A_{s}}{A_e},
\end{equation}
where $A_e$ is the single PE charge obtained from the PMT gain calibration, while $A_{s}$ is the fitting result from both Eq.~(\ref{waveformb}) and Eq.~(\ref{waveformt}).

The total number of scintillation photons $N_s$ can be obtained from $D_{s}$ divided by the detection efficiency $\varepsilon_{s}$,
\begin{equation}
N_s=\frac{D_{s}}{\varepsilon_{s}}=\frac{A_s}{\varepsilon_{s}\cdot A_e}.
\end{equation}
The detection efficiency was estimated with Geant4~\cite{G41, G42}-based Monte-Carlo simulation.
The geometry represented in Fig.~\ref{front} was implemented in the Geant4 simulation.
Muons were sampled according to the Gaisser formula~\cite{Gaisser}.
The simulation considered the emission spectra of different samples, quantum efficiency spectrum of PMT and attenuation length dependence on wavelength. Standard electromagnetic and muon-nucleus processes were both included.
More details of the simulation can be found in~\cite{li2016separation}.
The dominant uncertainty was found to be associated with the PMT quantum efficiency and estimated to be 10\%.

The scintillation light yield $Y$ can be calculated by
\begin{equation}
Y=\frac{N_s}{E_{\textrm{vis}}}=\frac{A_s}{\varepsilon_{s}\cdot A_e\cdot E_{\textrm{vis}}},
\label{ly}
\end{equation}
where $E_{\textrm{vis}}$ is the total visible energy and estimated to be $(69.1\pm1.9)$ MeV from the simulation.
It is important to note that this light yield includes the contribution from the hard UV portion of Cherenkov light because of the absorption and re-emission of photons.
Since these photons lost the directional information of the original Cherenkov photons, they were treated as part of the effective scintillation yield.

The quenching effect of muon is described by Birks' constant, which is 0.015\,cm/MeV for low-energy electrons~\cite{Birksf}.
The difference with and without Birks' constant results in an uncertainty of 2.8\%.

The distance from the light production point along the muon track in the scintillator to each PMT photocathode is several tens of centimeters, which is much shorter than the attenuation length at wavelength longer than 400\,nm.
In the shorter wavelength region, the attenuation length cannot be ignored. The attenuation length spectrum can be represented as the combination of all the solution components,
\begin{equation}
\frac{1}{L}=\sum\frac{n_i}{n_{0i}L_i},
\label{attcom}
\end{equation}
where $L$ is the total attenuation length, $n_i$ is the concentration of the $i$-th component and $L_i$ is the attenuation length measured at concentration $n_{0i}$.
\cite{Johnny} and \cite{Xiao} give the attenuation length spectra of pure LAB, LAB+3\,g/L PPO, LAB+3\,g/L PPO+15\,mg/L bis-MSB, and the inherent attenuation length spectra of PPO and bis-MSB can then be extracted from that of the compound according to Eq.~(\ref{attcom}),
\begin{equation}
\frac{1}{L_{\rm PPO}}=\frac{1}{L_{\rm LAB+PPO}}-\frac{1}{L_{\rm LAB}},
\end{equation}
\begin{equation}
\frac{1}{L_{\rm bis-MSB}}=\frac{1}{L_{\rm LAB+PPO+bis-MSB}}-\frac{1}{L_{\rm LAB}}-\frac{1}{L_{\rm PPO}}.
\end{equation}

For the LAB sample with 0.07\,g/L PPO and 13\,mg/L bis-MSB, the numbers of measured PEs at the top and bottom PMTs are shown in Table.~\ref{nPE}, and the uncertainties are all fitting errors.
The number of detected Cherenkov PEs was $5.47\pm0.22$.
The light yield for the sample was estimated to be $(4.01\pm0.60)\times10^3$\,photons/MeV.

\begin{table}[htbp]
\centering
\begin{tabular}{cccc}
\hline
& Top  & Bottom  & \\
\hline
Cherenkov     &  $--$ & $5.47\pm0.22$ \\
Scintillation &  $56.1\pm1.2$ & $56.1\pm1.2$ \\
\hline
\end{tabular}
\caption{Measured photoelectrons for the LAB sample with 0.07\,g/L PPO and 13\,mg/L bis-MSB. The scintillation light was assumed to be isotropic, and thereby gave the same photoelectrons for both the top and bottom PMTs.}
\label{nPE}
\end{table}

\subsection{Scanning of light yield and scintillation time}
We changed the concentration of PPO with different bis-MSB solutions, by measuring the scintillation light yield, rise time constant, decay time constant and Cherenkov photoelectron yield for each sample.
The results and the scintillation photon detection efficiencies are shown in Table~\ref{tab:tvy}.
These quantities can affect the performance of separation between scintillation and Cherenkov lights.

\begin{table}
\centering
\begin{tabular}{ccccccc}
\hline
\multirow{2}{*}{PPO} & \multirow{2}{*}{bis-MSB} & Detection  & Scintillation  & Rise time & Decay time & Number of  \\
 &   & efficiency  & light yield & constant  & constant  & Cherenkov \\
(g/L)&  (mg/L)     & ($\times10^{-4}$) & ($10^3$~photons/MeV)& (ns) &(ns) & photoelectrons\\
\hline
0    &   0  & $1.54\pm0.23$ & $2.53\pm0.38$ & $12.20\pm1.39$& $35.42\pm1.18$ & $10.06\pm0.40$\\	
0.02 &   0  & $2.23\pm0.33$ & $2.69\pm0.40$ & $3.72\pm0.10$ & $34.31\pm0.42$ & $9.08\pm0.36$ \\	
0.04 & 0.13 & $2.23\pm0.33$ & $3.04\pm0.46$ & $2.33\pm0.20$ & $30.55\pm0.30$ & $7.56\pm0.30$ \\	
0.07 &   0  & $2.20\pm0.33$ & $3.36\pm0.50$ & $1.33\pm0.13$ & $26.72\pm0.20$ & $8.58\pm0.34$ \\	
0.07 &   13 & $2.02\pm0.30$ & $4.01\pm0.60$ & $1.16\pm0.11$ & $26.76\pm0.19$ & $5.47\pm0.22$ \\	
0.1  &   0  & $2.18\pm0.33$ & $4.18\pm0.63$ & $1.15\pm0.10$ & $24.85\pm0.14$ & $8.26\pm0.33$ \\	
0.1  &  130 & $2.05\pm0.31$ & $5.45\pm0.82$ & $1.07\pm0.18$ & $20.94\pm0.10$ & $4.84\pm0.19$ \\	
0.5  &   0  & $2.13\pm0.32$ & $6.98\pm1.05$ & $1.00\pm0.05$ & $14.96\pm0.04$ & $7.55\pm0.30$ \\	
2    &   0  & $2.14\pm0.32$ & $9.57\pm1.44$ & $0.81\pm0.04$ & $8.88\pm0.02$  & $7.18\pm0.29$ \\	
3    &   0  & $2.10\pm0.32$ & $9.54\pm1.43$ & $0.64\pm0.04$ & $7.72\pm0.02$  & $7.24\pm0.29$ \\	
3    &  10  & $2.17\pm0.33$ & $11.59\pm1.74$& $0.77\pm0.04$ & $7.63\pm0.01$  & $5.08\pm0.20$ \\	
\hline
\end{tabular}
\caption{The scintillation photon detection efficiencies, scintillation light yields, time constants and Cherenkov PE numbers of the LAB samples with different concentrations of PPO and bis-MSB.\label{tab:tvy}}
\end{table}

\begin{figure}
\centering
\includegraphics[width=0.7\textwidth]{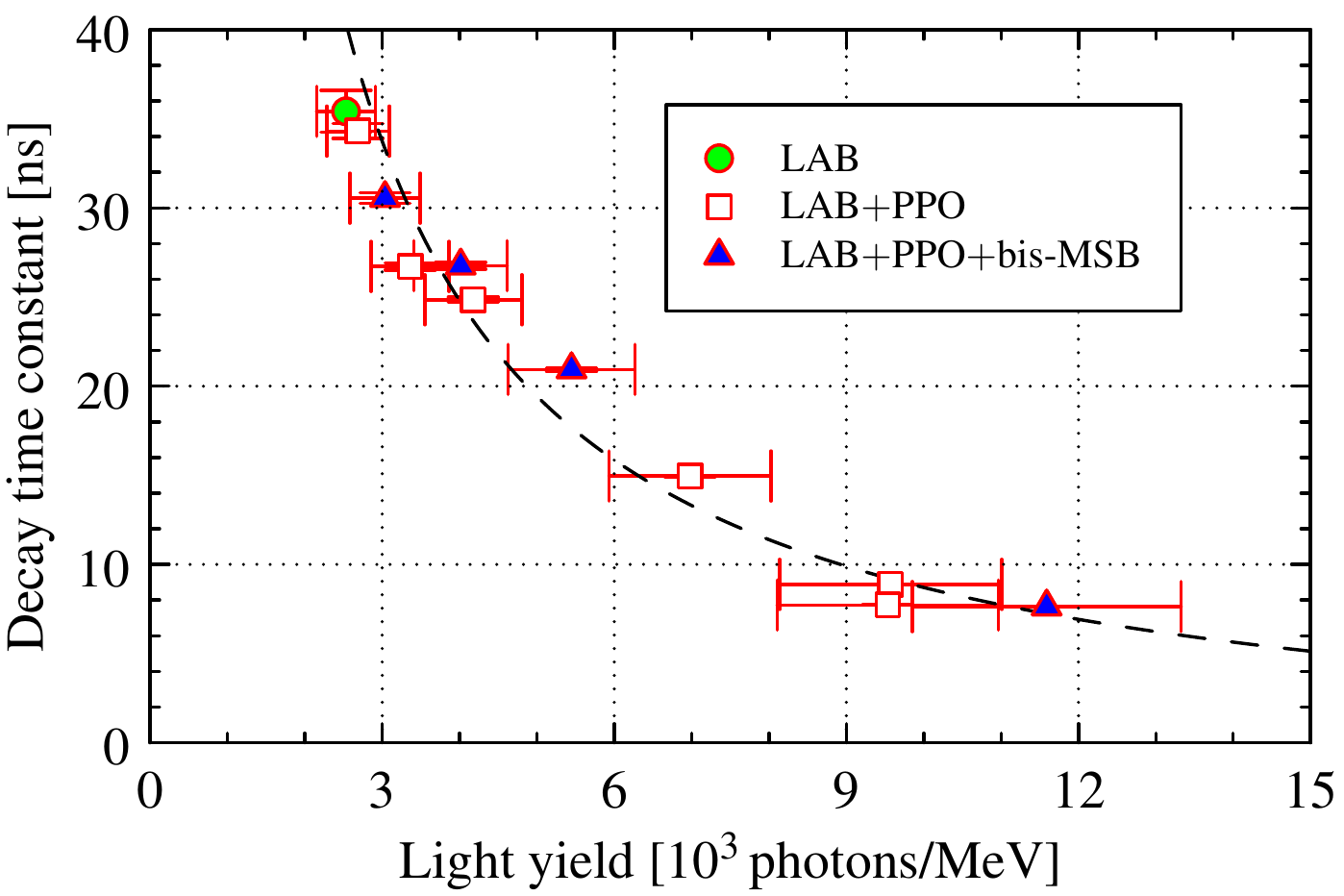}
\caption{Decay time constant versus scintillation light yield for different concentrations of LAB, PPO, and bis-MSB solutions. The dashed line is a fit of Eq.~(\ref{TT}).}
\label{fig:relationship}
\end{figure}

The decay time constants and scintillation light yields are plotted in Fig.~\ref{fig:relationship} for all the test samples and showing an inverse relationship. The effect of wavelength shifter bis-MSB on decay time constants and scintillation light yields is relatively insignificant at low concentrations.
Increasing PPO concentration will result in higher light yields and smaller time constants.

\begin{figure}
\centering
\includegraphics[width=0.7\textwidth]{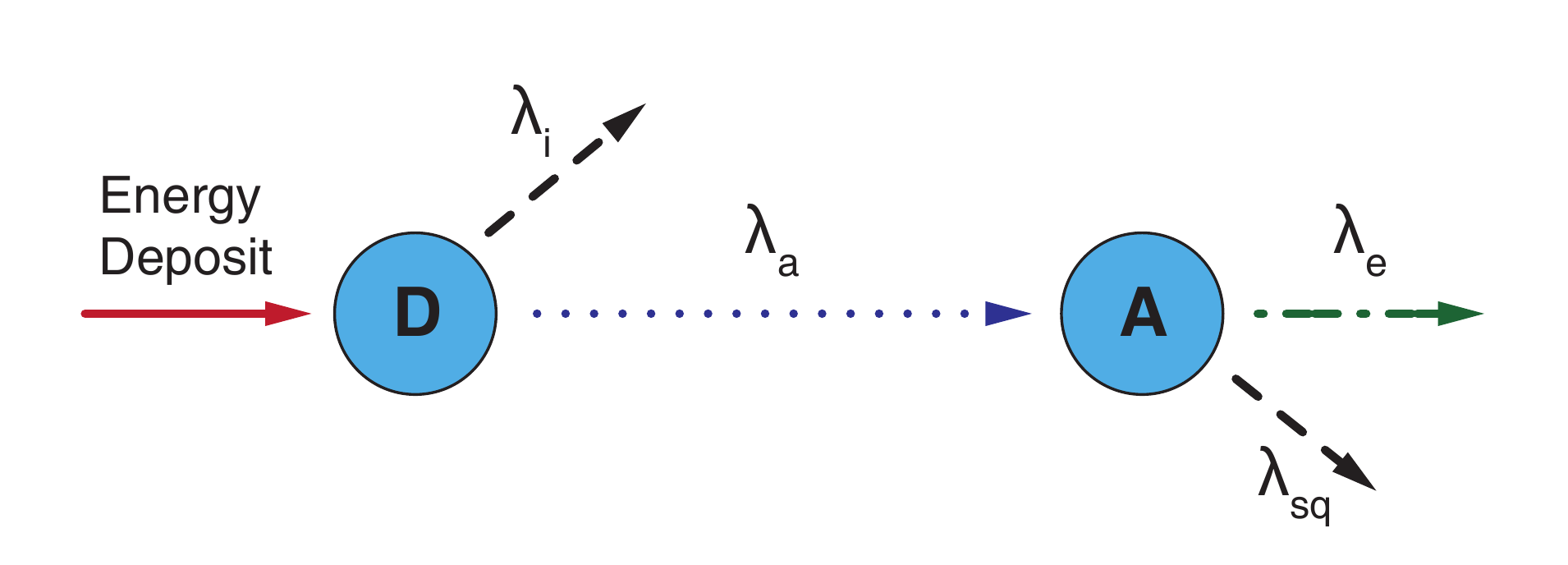}
\caption{A simplified model of energy transfer paths between donor (D) and acceptor (A) molecules in a liquid scintillator.~\cite{LYAR}}
\label{fig:EnergyTransfer}
\end{figure}

To understand this inverse relationship between the scintillation light yield and the decay time constant, the mechanism of light emission in the scintillator was examined. As shown in Fig.~\ref{fig:EnergyTransfer}, incident charged particles in a liquid scintillator deposit their energies, some of which can be transferred between molecules.

The light yield $Y$ from the energy transfer was modeled by \cite{LYAR}, as expressed by Eq.~(\ref{eq:LYAR}), in which the amounts of energy transfer is represented by the PPO concentration $A$,
\begin{equation}
\label{eq:LYAR}
Y=D\cdot\frac{1}{1+\frac{\lambda_{sq}A}{\lambda_e}}\cdot\frac{1}{1+\frac{\lambda_i}{\lambda_aA}}
\end{equation}
where $D$ is the number of excited solvent molecules, $\lambda_{sq}$ is the self-quenching factor, $\lambda_e$ is the rate of photon emission after self-quenching, $\lambda_i$ is the internal loss factor of solvent molecules, and $\lambda_a$ is the energy transfer from the solvent (donor) molecules to the solute (acceptor) PPO molecules.

The self-quenching effect is due to the interaction between unexcited and excited PPO molecules. The excitation energy is lost by collision and since the self-quenching can be neglected for the case of low PPO concentration ($<10$\,g/L), the light yield can thus be simplified as
\begin{equation}
\label{LYA}
Y=\frac{DA}{A+\frac{\lambda_i}{\lambda_a}}
\end{equation}

The decay time constant $\tau$ can be described as the sum of the solute intrinsic lifetime $\tau_s$ and energy migration transfer (or \textit{hopping}) time~\cite{TAR},
\begin{equation}
\label{TA}
\tau=\tau_s+\frac{A_0}{k_hA}
\end{equation}
where $k_h$ is the effective energy migration transfer rate for a given concentration $A_0$. The number of energy migration transfer processes caused by solvent-solvent collisions is inverse proportional to the PPO concentration $A$.

Combining Eqs.~(\ref{LYA}) and (\ref{TA}), we can obtain the relationship between the light yield and decay time constant,
\begin{equation}
\label{TT}
\tau=\tau_s-\frac{A_0\lambda_a}{k_h\lambda_i}+\frac{A_0\lambda_aD}{k_h\lambda_i}\cdot\frac{1}{Y}\equiv \tau_0+\frac{C}{Y}
\end{equation}
where $\tau_0\equiv\tau_s-\frac{A_0\lambda_a}{k_h\lambda_i}$ and $C\equiv\frac{A_0\lambda_aD}{k_h\lambda_i}$.
This clearly indicates an inverse relationship between the decay time constant $\tau$ and the scintillation light yield $Y$.
We observed that Eq.~(\ref{TT}) was consistent with our measurements as indicated in Fig.~\ref{fig:relationship}.

As shown in Table~\ref{tab:tvy}, when the concentration increases beyond 0.1\,g/L, the decay time constant is so small that the separation between scintillation light and Cherenkov light from the pulse shape discrimination becomes difficult. In the mean time, the addition of bis-MSB reduces the number of Cherenkov photons , even though it can shift the wavelength to the detectable region.
Samples of LAB with 0.07$\sim$1\,g/L PPO and 0$\sim$13\,mg/L bis-MSB could be good slow liquid scintillator candidates and should be more thoroughly investigated.

\section{Emission spectrum \label{sec:emission}}

\begin{figure}
\centering
\includegraphics[width=0.7\textwidth]{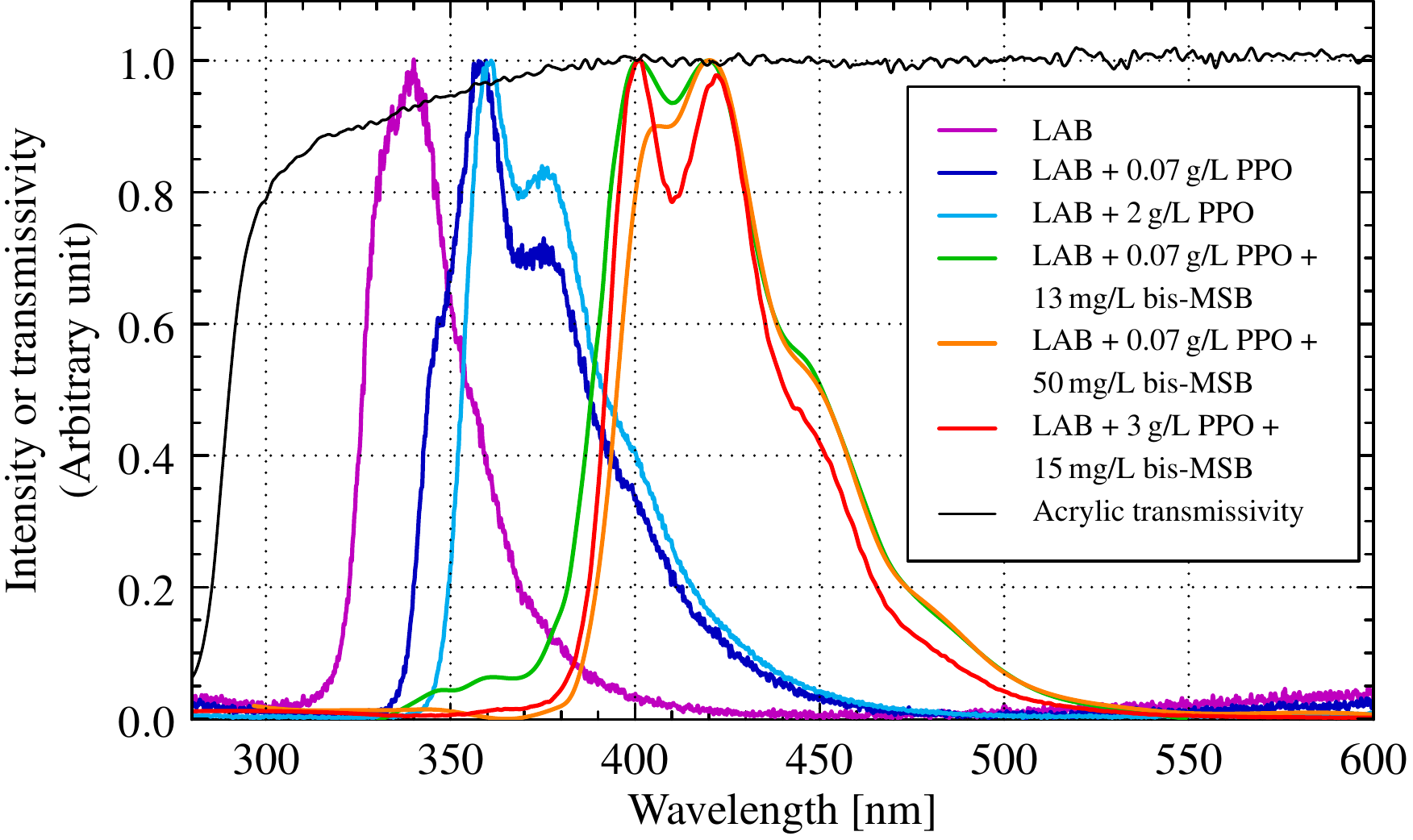}
\caption{Emission spectra of the pure LAB, usual LS (red line) and slow liquid scintillator candidates.\label{fig:emission}}
\end{figure}

The emission spectra of the candidate samples were measured using a RTI fluorescence spectrometer (made by Ocean Optics) excited at 260~nm.
The relevant spectra are shown in Fig.~\ref{fig:emission}.
LAB emits light at 280$\sim$300\,nm. However, in a bulk solution, both re-absorption and re-emission occur during the light propagation process and shift the wavelength upward. It is important to note that the wavelength range from 380 to 500\,nm is detectable to the PMT and transparent to the acrylic. From the emission spectra, it was possible to conclude that a significant amount of additional bis-MSB should be required. Fig.~\ref{fig:emission} shows that the additions of 13\,mg/L or more bis-MSB have similar wavelength spectra.
The formulas with no bis-MSB addition have a better Cherenkov separation capability, but may lead to a lower photoelectron yield. There is a trade-off between the photoelectron yield and Cherenkov separation capability. These emission spectra were implemented in the simulation for evaluating the detection efficiency in Section~\ref{sec:Yield-Time}.
It is important to note that the sample container cell of the RTI fluorescence spectrometer is $10\times10\times10$\,cm$^3$, and the spectra measured is not at the origin of excitation. This was reflected in our simulation for the 15.4 L device.

Combining the emission spectra with the scanning results in Table~\ref{tab:tvy}, we chose the formula of LAB with 0.07\,g/L of PPO and 13\,mg/L of bis-MSB as our slow liquid scintillator candidate, since  it has reasonable light yield and time constants, while maintaining more than half of the Cherenkov photons with respect to the pure LAB. The emission spectrum was shifted to the detectable range above 390\,nm, falling into the detectable region with almost no optical loss in acrylics (see Section~\ref{sec:acrylic}).

\section{Optical transmission of acrylic\label{sec:acrylic}}
Since acrylics are compatible with LAB-based liquid scintillators in terms of chemical and optical properties, they are widely used for the scintillator vessels in neutrino experiments.
However, to maintain a kiloton-scale liquid scintillator, the acrylic vessel should be at least several centimeters thick, as used in the SNO experiment~\cite{SNO}.
As a result, the optical transmission loss in acrylic cannot be ignored. To evaluate the effect, we performed a qualitative study on the transmission for a UV transparent acrylic sample (UV transparent type made by DONCHAMP, China).

\begin{figure}
\centering
\includegraphics[width=0.7\textwidth]{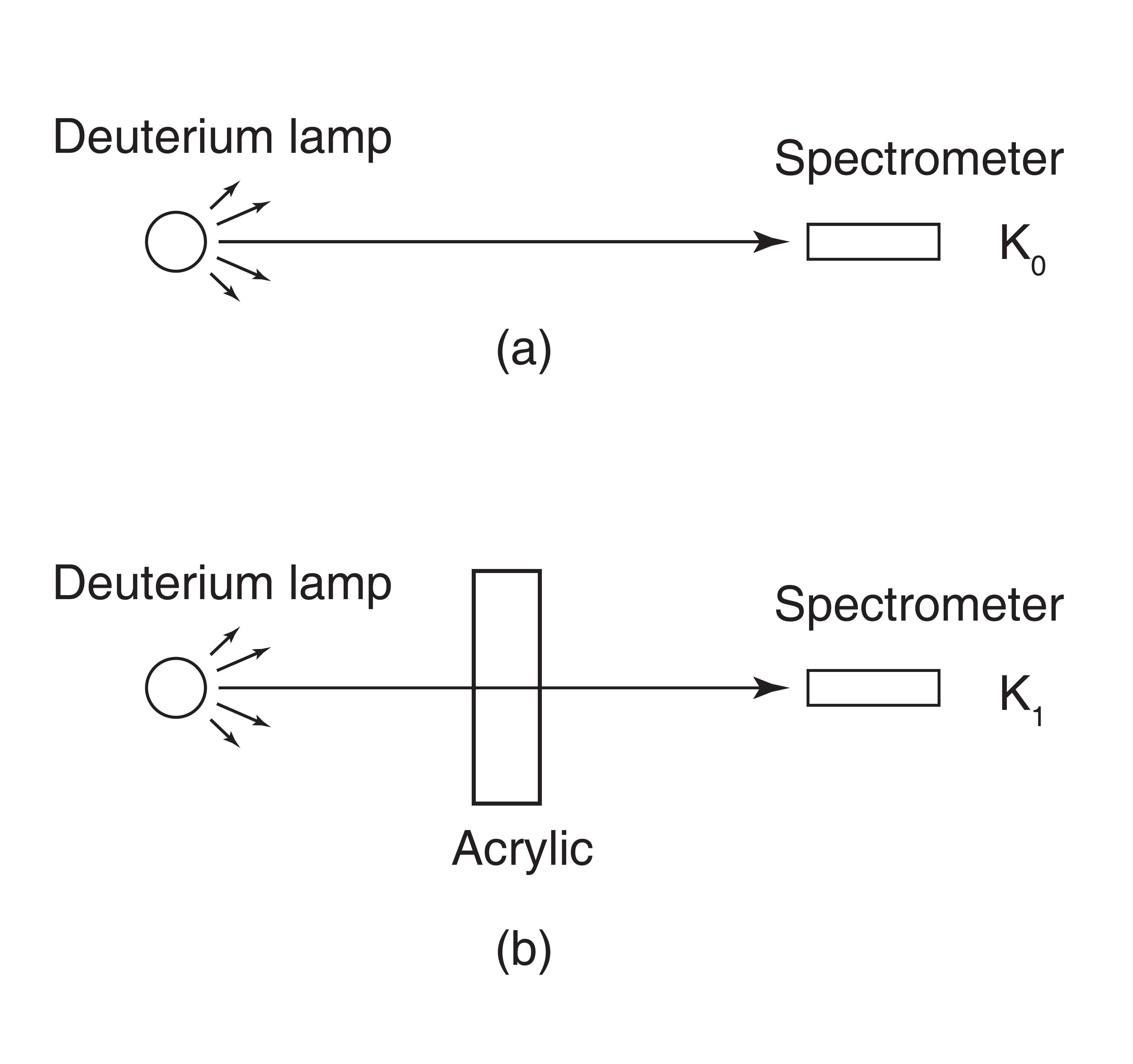}
\caption{The schematic of the apparatus used for the acrylic optical transmission measurement.  \label{fig:acrylicApp}}
\end{figure}

\begin{figure}
\centering
\includegraphics[width=0.7\textwidth]{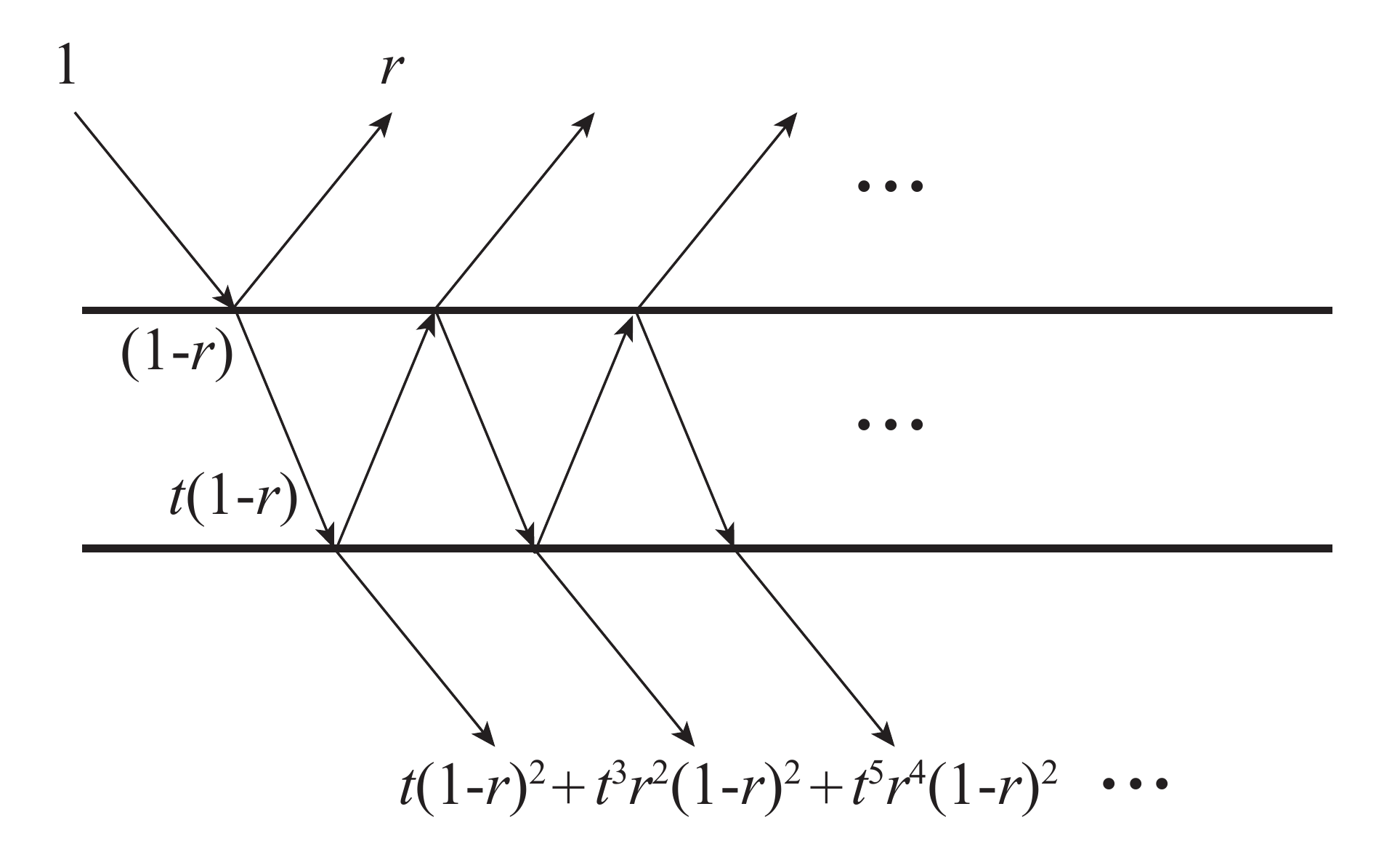}
\caption{Transmissivity and reflectivity of an acrylic layer. The intensity of the incident light beam is 1, the transmissivity of the sample is $t$, and the reflectivity in air is $r$. The light beam hits the interface at a perpendicular angle in the experiment, but in this figure the light beams are drawn at a non-perpendicular angle to easily distinguish the incident from reflection light beams.\label{fig:transmissivity}}
\end{figure}

The test stand included a deuterium lamp and a spectrometer (Ocean Optics) with a 10-mm-thick test sample plate in between, as shown in Fig.~\ref{fig:acrylicApp}. The lamp light was set perpendicularly incident to the acrylic plate and the spectrometer was used to measure the transmission light. For comparison, the light intensity spectra was measured. The corresponding light intensity spectra were referred to as $K_1$ and $K_0$. The ratio between $K_1$ and $K_0$ is a function of the transmissivity $t$ and  reflectivity $r$ for the acrylic sample as illustrated in Fig.~\ref{fig:transmissivity}.

For the case of a vertical incident light, the transmission intensity should be $t(1-r)^2+t^3r^2(1-r)^2+\cdots$, and the ratio can therefore be written as
\begin{equation}
\frac{K_1}{K_0}=\frac{t(1-r)^2}{1-t^2r^2}.
\label{eq:trans}
\end{equation}
The reflectivity $r$ can be derived from the Fresnel formula ,
\begin{equation}
r=\left(\frac{n-1}{n+1}\right)^2.
\label{eq:r}
\end{equation}
where $n$ is the reflective index. The curve of transmissivity as a function of the wavelength was obtained from Eqs.~(\ref{eq:trans}) and (\ref{eq:r}) is shown in Fig.~\ref{fig:emission}. The acrylic sample was found to be nearly transparent in the visible light wavelength range (i.e., $>400$\,nm) and became almost opaque in the wavelength range  below 270\,nm.
As shown in Fig.~\ref{fig:emission}, the spectrum of emission light for the pure LAB was below 400\,nm and should be shifted upward to avoid the absorption of acrylics.

\section{Attenuation length of the scintillator\label{sec:attenuation}}
\begin{figure}
\centering
\subfigure[\label{attApp}]{
\includegraphics[width=0.25\textwidth]{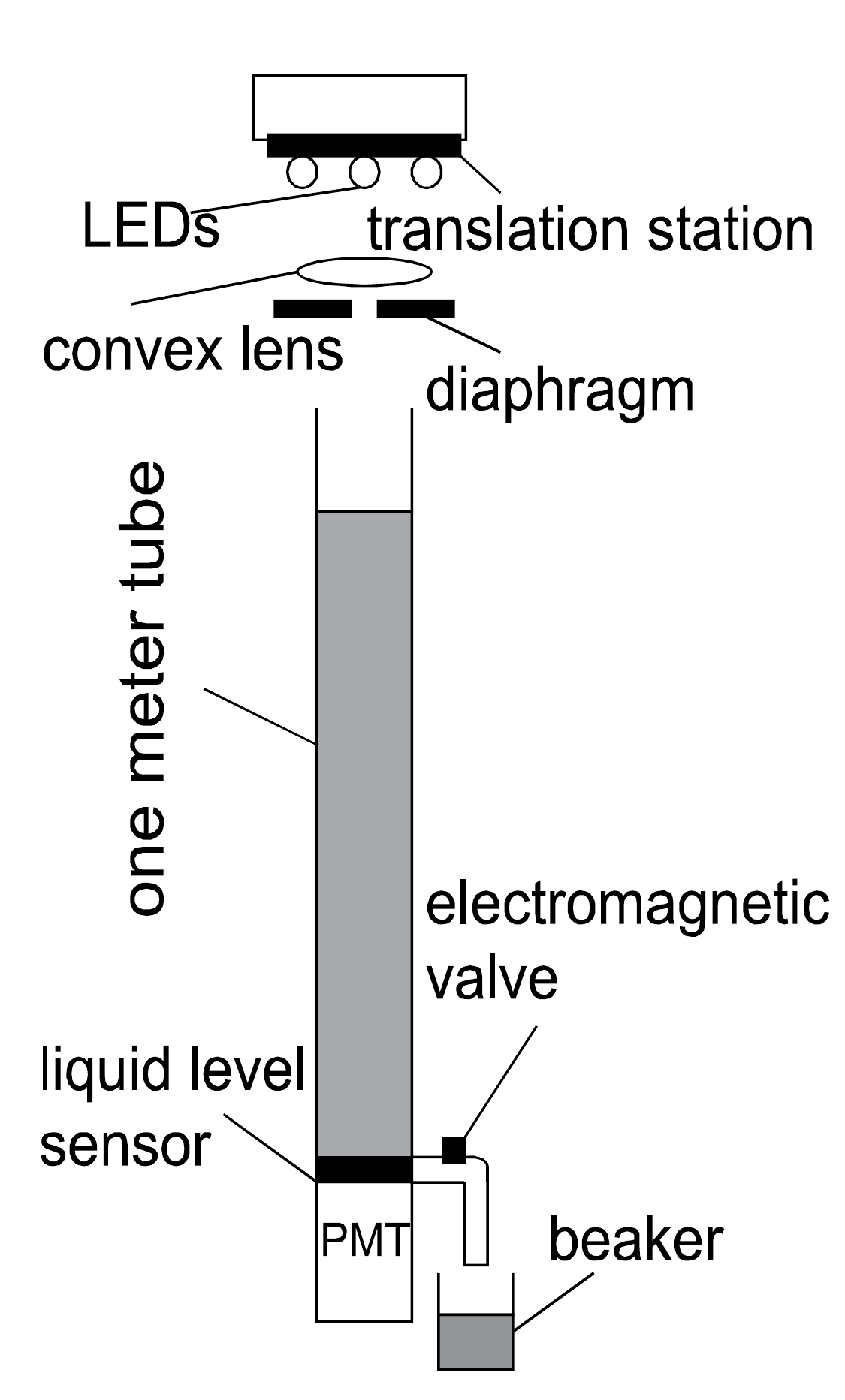}}
\hspace{0.05\textwidth}
\subfigure[\label{LEDSpe}]{
\includegraphics[width=0.45\textwidth]{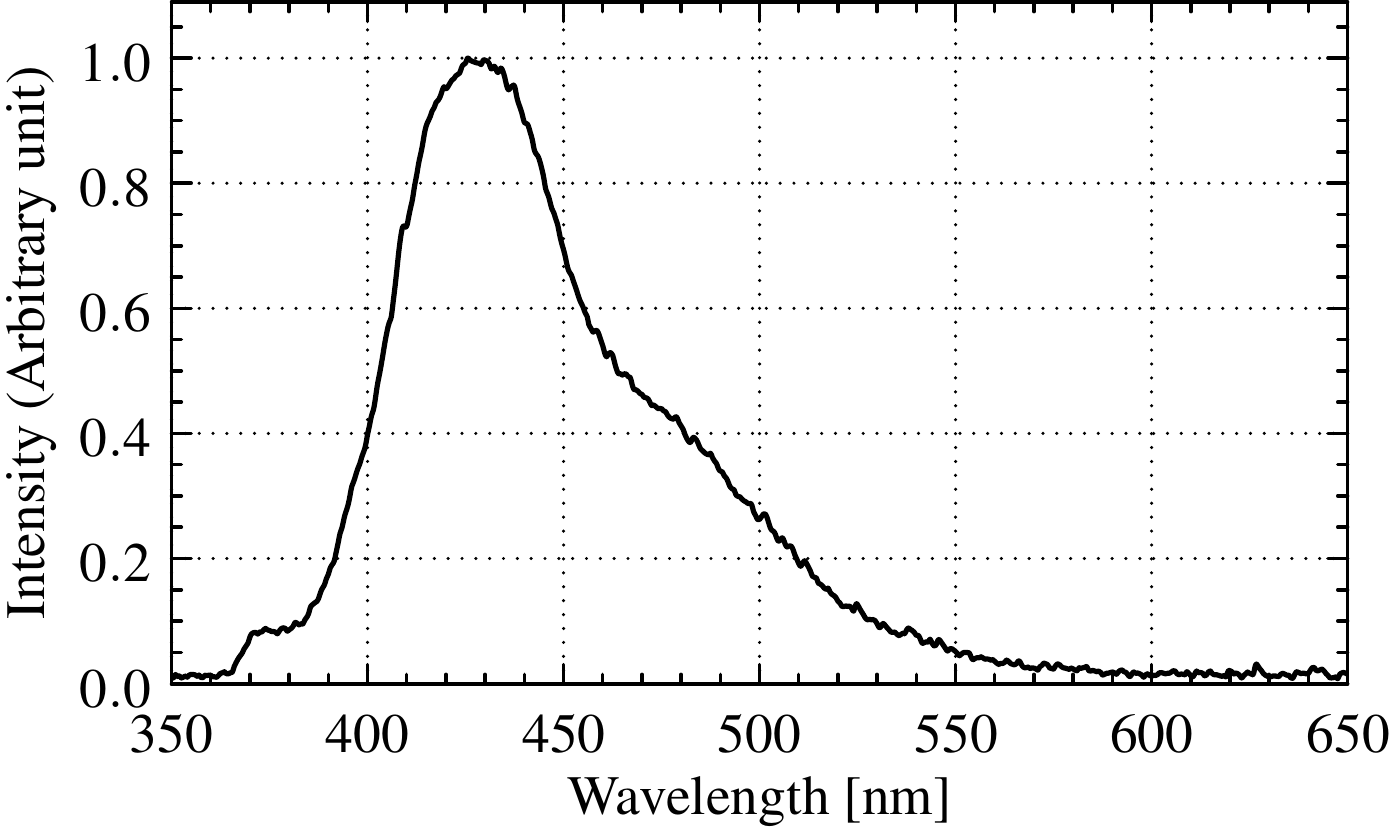}}
\caption{(a) A schematic of the variable pathlength photometer used for attenuation length measurement. (b) The emission spectrum of the LED mounted at the top. The central peak is 430\,nm.}
\end{figure}

If all the chemical components in the scintillator were precisely known, the attenuation length might have been easily obtained using Eq.~(\ref{attcom}). Given the fact that the impurity of the sample is difficult to know, a photometer was used to measure  the attenuation of the slow liquid scintillator candidates~\cite{Atten, Atten2}. A schematic of this photometer is shown in Fig.~\ref{attApp}.

An LED lamp was mounted at the top of the photometer. Light was refocused from a lens to ensure it to travel through a diaphragm and a 1\,m-long stainless steel pipe filled with the liquid scintillator. The liquid level in the pipe was controlled by a solenoid valve and a liquid level sensor. A PMT (Hamamatsu R7724, 51\,mm diameter) was installed at the bottom of the equipment to receive light, and the wavelength of the response displayed a maximum at 420\,nm~\cite{PMT}.

The slow liquid scintillator used in the measurement was the LAB with 0.07\,g/L of PPO and 13\,mg/L of bis-MSB, which had an emission spectrum that partially overlapped with that of the LED light used in the experiment.

\begin{figure}
\centering
\includegraphics[width=0.7\textwidth]{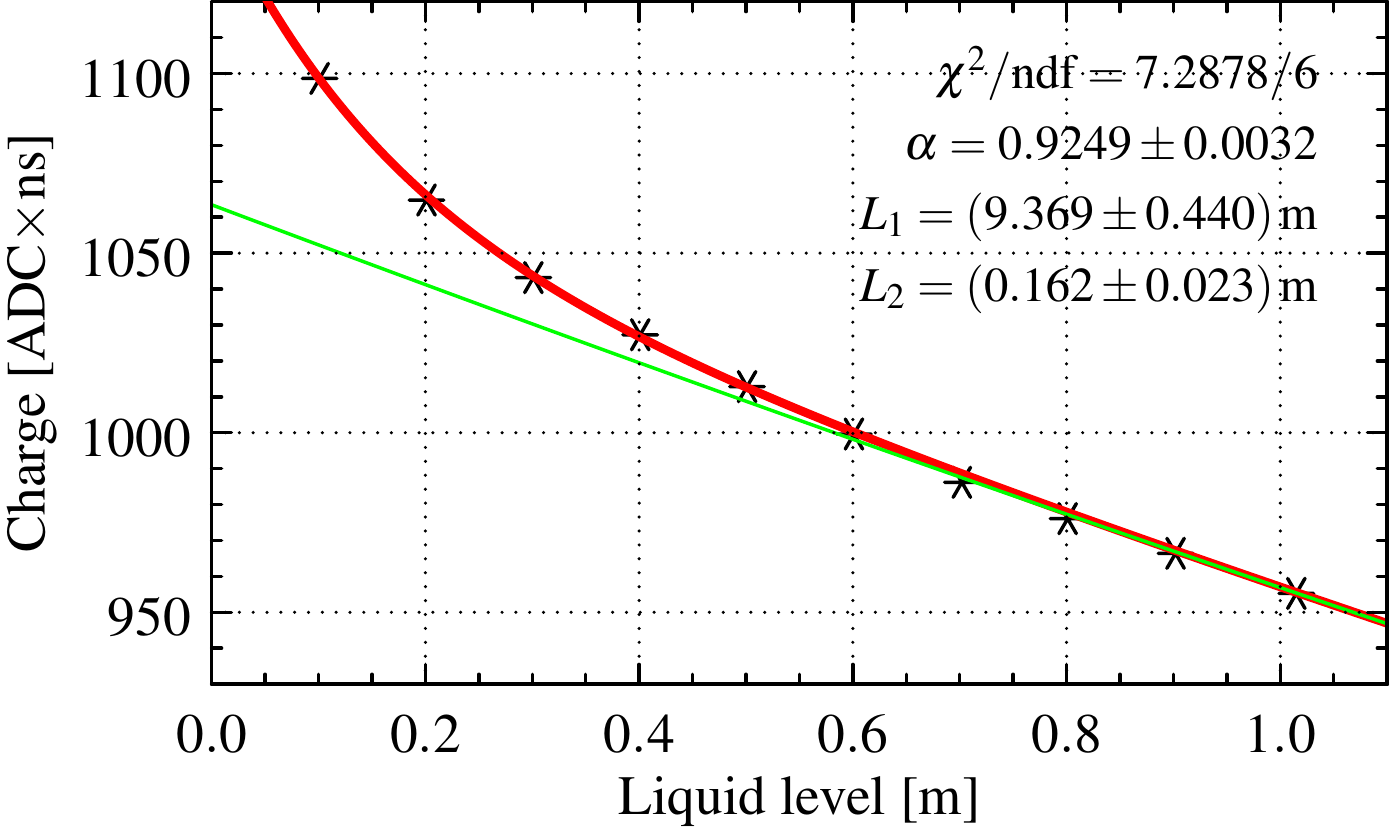}
\caption{Two-exponential fitting results of LAB with 0.07\,g/L PPO and 13\,mg/L bis-MSB. The green line reflects the long attenuation length component.
\label{LS}}
\end{figure}

As shown in Fig.~\ref{LEDSpe}, the LED spectrum is not monochromatic, so the light attenuation cannot be described by a simple exponentially decreasing curve. Instead, the intensity of transmission light $I(x)$ is described by a weighted average of the LED spectrum $f(\lambda)$,
\begin{equation}
I(x)=I_0\int f(\lambda)\ue^{-x/L(\lambda)}\ud\lambda,
\end{equation}
where $I_0$ is the intensity of incident light. For the sake of convenience, two exponentials were used to describe the data.
\begin{equation}
I(x)=I_0\left[\alpha\ue^{-x/L_1}+(1-\alpha)\ue^{-x/L_2}\right],
\end{equation}
where $\alpha$ is the fraction of the component with a longer attenuation length (referred to as $L_1$), while $L_2$ is the shorter attenuation length. The fitting result is shown in Fig.~\ref{LS}, in which $\alpha$ was determined to be $0.925\pm0.003$, $L_1$ and $L_2$ were determined to be $(9.37\pm0.44)$\,m and $(0.16\pm0.02)$\,m, respectively.

It should be noted that the measured attenuation lengths actually included the contribution from the absorption, re-emission, and scattering effects~\cite{Zhouxiang1,Zhouxiang2}. It is understood that with a purification process the attenuation length can be extended to 20\,m.

\section{Performance evaluation for the candidate sample\label{sec:Candidate}}
In this section, the possibility of using the candidate samples to satisfy both the detection requirements of solar and supernova relic neutrino experiments~\cite{wei2017discovery} is proposed. The detection feature is demonstrated by an analytical calculation with an empirical detector model.

\subsection{Solar neutrino study}
\begin{figure}
\centering
\includegraphics[width=0.7\textwidth]{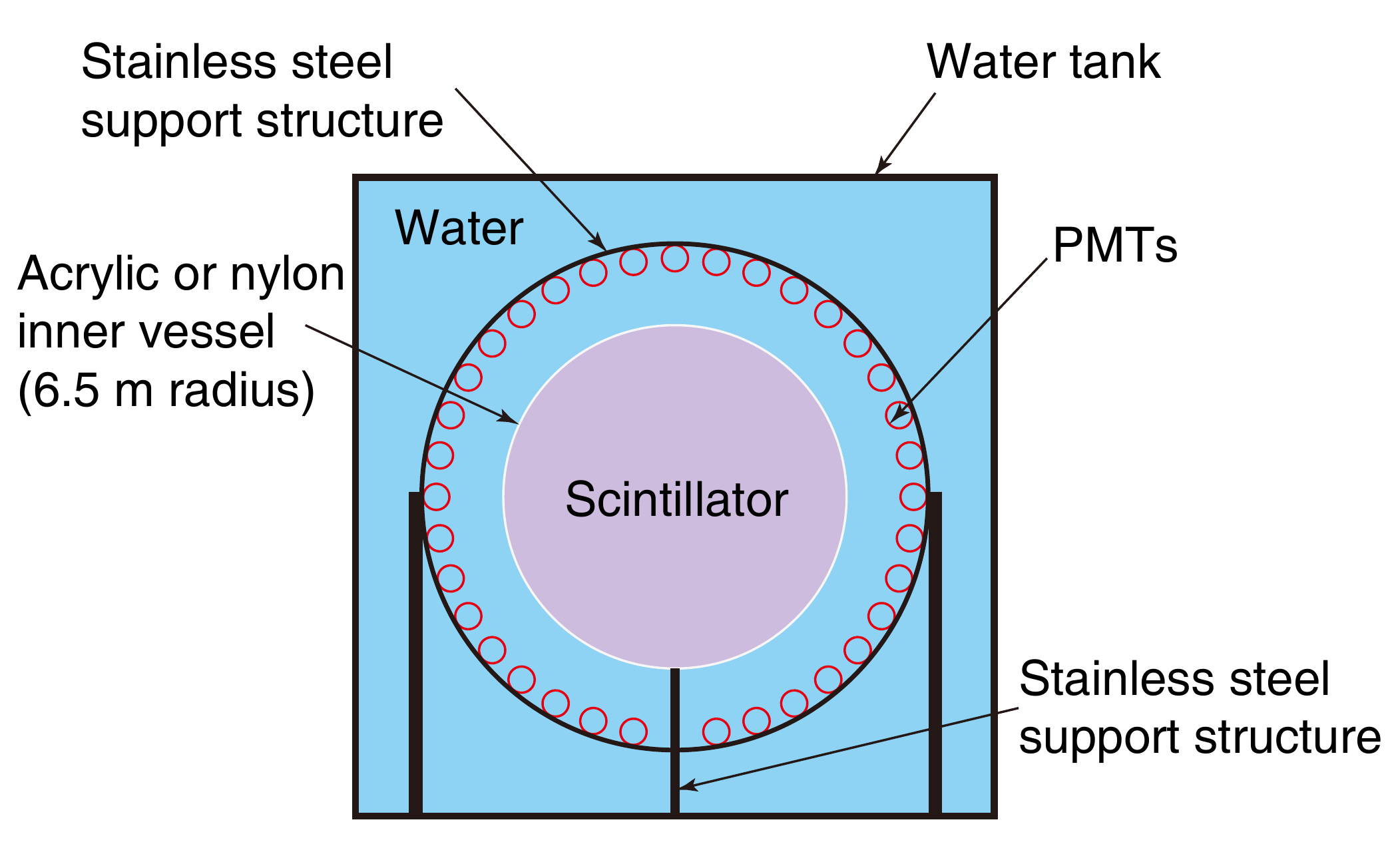}
\caption{A typical solar neutrino observatory with 1 kiloton target mass.
\label{1kt}}
\end{figure}

The impact of the slow liquid scintillator on future solar neutrino experiments was studied, for example the neutrino experiment at Jinping~\cite{beacom2017physics}, in which a minimum light yield of 500\,photoelectrons/MeV is required. To evaluate the effect, we set up a detector model, as shown in Fig.~\ref{1kt}, which was similar to SNO+~\cite{SNOplus}. The target material filled in an acrylic or nylon inner vessel was 1 kiloton (6.5\,m radius) liquid scintillator, 4,000 or more PMTs were placed around the inner vessel in the buffer water, the height and diameter of the water tank were both 12\,m.

The photoelectrons yield $E$ of a scintillator neutrino detector is calculated using
\begin{equation}
E=Y\cdot \epsilon \cdot c \cdot q
\end{equation}
where $Y$ is the light yield, $\epsilon$ is the light propagation efficiency, $c$ is the photocathode coverage and $q$ is the quantum efficiency.

We assumed the PMT photocathode coverage and average quantum efficiency to be 70\% and 20\%, respectively. Efficiency loss due to the acrylic vessel was ignored owing to the wavelength shifter. For the effect of self attenuation, the light propagation efficiency was 46.2\% in the center of the detector.
For the sample of LAB with 0.07\,g/L of PPO and 13\,mg/L of bis-MSB, the scintillation light yield is roughly $E=4010\times46.2\%\times70\%\times20\%=260\,{\rm PE/MeV}$.

If the PMT photocathode coverage can reach $100\%$ with the help of light concentrators~\cite{zhiyu}, PMTs with high quantum efficiency ($>30\%$) can be adopted, and the attenuation length can reach up to 15\,m~\cite{15m} (i.e., the average propagation efficiency was $60\%$, then the photoelectrons yield could be increased to $720\,\rm{PE/MeV}$, or $3.7\%$ at 1\,MeV of detectable energy). This value meets the requirement in the Jinping proposal for the solar neutrino study~\cite{beacom2017physics}.

\subsection{Supernova relic neutrino study}
The detection of supernova relic neutrinos~\cite{wei2017discovery} does not have a stringent requirement for the light yield.
The suppression of the neutral and charged current backgrounds induced by atmospheric neutrinos relies on the identification of neutrons and protons. For the region of interest, the positron signals are between 10$\sim$30\,MeV.
Unlike the recoiled protons, these positrons are over Cherenkov production threshold.
About 500 Cherenkov photons are emitted in the 400$\sim$700\,nm range for a 10\,MeV positron. Considering a quantum efficiency of 30\%, a light propagation efficiency of 60\% and photocathode coverage of 100\%, approximately 90 photoelectrons are predicted to be detected by the PMTs in the forward Cherenkov ring direction.
Based on the separation capability shown in Fig.~\ref{fig:fit}, where the signal to background ratio (Cherenkov to scintillation) within the first 10\,ns is approximately 1. The 90 photoelectrons from the Cherenkov light should have been easily identified.
The samples with 0.07\,g/L of PPO and 13\,mg/L of bis-MSB or formulas with similar PPO and bis-MSB concentrations could be a potential candidate to supernova relic neutrino detection and further studies are necessary to assess the gain in sensitivity.

\section{Conclusion and outlook\label{sec:conclusion}}

In this study, the liquid scintillator mixtures of LAB, PPO and bis-MSB solution with different compounding ratios were investigated.
An inverse relationship between the light yields and decay time constants for these samples was observed. The relationship was understood by the mechanism of the energy transfer between scintillator molecules. The emission spectra of these samples were also reported. The addition of PPO and bis-MSB could enhance the light yield and shift the emission spectrum toward a more optically transparent region.

For the first time, samples of LAB with around 0.07\,g/L of PPO and 13\,mg/L of bis-MSB or formulas with similar concentrations are shown to display a good balance between scintillation decay time and light yield. This combination with the PMT detection approach could serve as good slow liquid scintillator candidates for solar and supernova relic neutrino experiments. The concentrations of PPO and bis-MSB could be further optimized by using a large test apparatus~\cite{1ton} and by performing extensive offline simulations and analyses.

For the direction reconstruction performance of low-energy electrons below 10\,MeV, a more complicated full detector simulation and reconstruction method is being studied. Its impact on the detection of geoneutrinos through the electron-neutrino scattering will be reported.

\section*{Acknowledgements}
This work was supported in part by, the National Natural Science Foundation of China (No.~11620101004 and 11475093), the Key Laboratory of Particle \& Radiation Imaging (Tsinghua University), and the CAS Center for Excellence in Particle Physics (CCEPP), and portion of this work performed at Brookhaven National Laboratory is supported in part by the United States Department of Energy under contract DE-AC02-98CH10886.

\section*{References}

\bibliography{main}

\begin{thebibliography}{10}
\expandafter\ifx\csname url\endcsname\relax
  \def\url#1{\texttt{#1}}\fi
\expandafter\ifx\csname urlprefix\endcsname\relax\def\urlprefix{URL }\fi
\expandafter\ifx\csname href\endcsname\relax
  \def\href#1#2{#2} \def\path#1{#1}\fi

\bibitem{kang2010status}
K.-J. Kang, et~al., Status and prospects of a deep underground laboratory in
  china, in: Journal of Physics: Conference Series, Vol. 203, IOP Publishing,
  2010, p. 012028.

\bibitem{beacom2017physics}
J.~F. Beacom, et~al., Physics prospects of the jinping neutrino experiment,
  Chinese physics C 41~(2) (2017) 023002.

\bibitem{geo-WanLinyan}
L.~Wan, G.~Hussain, Z.~Wang, S.~Chen, Geoneutrinos at jinping: Flux prediction
  and oscillation analysis, Phys. Rev. D 95 (2017) 053001.

\bibitem{geo-Bill}
O.~$\check{S}$r$\acute{a}$mek, et~al., Revealing the earth��s mantle from
  the tallest mountains using the jinping neutrino experiment, Phys. Rev. D 6
  (2016) 33034.

\bibitem{alonso2014advanced}
J.~Alonso, et~al., Advanced scintillator detector concept (asdc): A concept
  paper on the physics potential of water-based liquid scintillator, arXiv
  preprint arXiv:1409.5864.

\bibitem{YEH201151}
M.~Yeh, et~al., A new water-based liquid scintillator and potential
  applications, Nucl. Instrum. Methods A 660~(1) (2011) 51 -- 56.

\bibitem{wei2017discovery}
H.~Wei, Z.~Wang, S.~Chen, Discovery potential for supernova relic neutrinos
  with slow liquid scintillator detectors, Physics Letters B 769 (2017)
  255--261.

\bibitem{JINST}
C.~Aberle, et~al., Measuring directionality in double-beta decay and neutrino
  interactions with kiloton-scale scintillation detectors, JINST 9 (2014)
  P06012.

\bibitem{Fukuda:2016yjg}
Y.~Fukuda, {ZICOS - New project for neutrinoless double beta decay experiment
  using zirconium complex in liquid scintillator}, J. Phys. Conf. Ser. 718~(6)
  (2016) 062019.

\bibitem{elagin2017separating}
A.~Elagin, et~al., Separating double-beta decay events from solar neutrino
  interactions in a kiloton-scale liquid scintillator detector by fast timing,
  Nucl. Instrum. Methods A 849 (2017) 102--111.

\bibitem{ciuffoli2016neutrino}
E.~Ciuffoli, J.~Evslin, F.~Zhao, Neutrino physics with accelerator driven
  subcritical reactors, Journal of High Energy Physics 2016~(1) (2016) 4.

\bibitem{ciuffoli2014leptonic}
E.~Ciuffoli, J.~Evslin, X.~Zhang, The leptonic cp phase from muon decay at rest
  with two detectors, Journal of High Energy Physics 2014~(12) (2014) 51.

\bibitem{geoZhe}
Z.~Wang, S.~Chen, Reveal the mantle and k-40 components of geoneutrinos with
  liquid scintillator cherenkov neutrino detectors, arXiv preprint
  arXiv:1709.03743.

\bibitem{LSND}
R.~A. Reeder, et~al., Dilute scintillators for large volume tracking detectors,
  Nuclear Instruments and Methods in Physics Research A 353 (1993) 353--366.

\bibitem{winn1985water}
D.~R. Winn, D.~Raftery, Water-based scintillators for large-scale liquid
  calorimetry, IEEE Transactions on Nuclear Science 32~(1) (1985) 727--732.

\bibitem{bignell2015characterization}
L.~J. Bignell, et~al., Characterization and modeling of a water-based liquid
  scintillator, Journal of Instrumentation 10~(12) (2015) P12009.

\bibitem{bignell2015measurement}
L.~J. Bignell, et~al., Measurement of radiation damage of water-based liquid
  scintillator and liquid scintillator, Journal of Instrumentation 10~(10)
  (2015) P10027.

\bibitem{So:2014hua}
S.~H. So, et~al., {Development of a Liquid Scintillator Using Water for a Next
  Generation Neutrino Experiment}, Adv. High Energy Phys. 2014 (2014) 327184.

\bibitem{li2016separation}
M.~Li, et~al., Separation of scintillation and cherenkov lights in linear alkyl
  benzene, Nucl. Instrum. Methods A 830 (2016) 303--308.

\bibitem{CHESS}
J.~Caravaca, et~al., Experiment to demonstrate separation of cherenkov and
  scintillation signals, Phys. Rev. C 95 (2017) 055801.

\bibitem{CHESS2}
J.~Caravaca, et~al., Cherenkov and scintillation light separation in organic
  liquid scintillators, arXiv preprint arXiv:1610.02011.

\bibitem{LAPPD}
B.~Adams, et~al., Measurements of the gain, time resolution, and spatial
  resolution of a 20$\times$20 cm$^2$ mcp-based picosecond photo-detector,
  Nuclear Instruments and Methods in Physics Research A 732 (2013) 392�C396.

\bibitem{DYB-LS1}
M.~Yeh, A.~Garnov, R.~Hahn, Gadolinium-loaded liquid scintillator for
  high-precision measurements of antineutrino oscillations and the mixing
  angle, $\theta_{13}$, Nuclear Instruments and Methods in Physics Research A
  578~(1) (2007) 329 -- 339.

\bibitem{DYB-LS2}
Y.~Ding, et~al., A new gadolinium-loaded liquid scintillator for reactor
  neutrino detection, Nuclear Instruments and Methods in Physics Research A
  584~(1) (2008) 238 -- 243.

\bibitem{PMT}
Hamamatsu photonics,
  \url{https://www.hamamatsu.com/resources/pdf/etd/R1828-01_R2059_TPMH1259E.pdf}.

\bibitem{Johnny}
J.~Goett, et~al., Optical attenuation measurements in metal-loaded liquid
  scintillators with a long-pathlength photometer, Nuclear Instruments and
  Methods in Physics Research Section A: Accelerators, Spectrometers, Detectors
  and Associated Equipment 637 (2011) 47.

\bibitem{Birks}
J.~B. Birks, The Theory and Practice of Scintillation Counting, Pergamon, 1964.

\bibitem{G41}
S.~Agostinelli, et~al., Geant4 - a simulation toolkit, Nucl. Instrum. Methods A
  506~(3) (2003) 250 -- 303.

\bibitem{G42}
J.~Allison, et~al., Geant4 developments and applications, IEEE Transactions on
  Nuclear Science 53~(1) (2006) 270--278.

\bibitem{Gaisser}
T.~K. Gaisser, Cosmic Rays and Particle Physics, Cambridge University Press,
  1990.

\bibitem{Birksf}
C.~Aberle, et~al., Light output of double chooz scintillators for low energy
  electrons, Journal of Instrumentation 6 (2011) P11006.

\bibitem{Xiao}
H.-L. Xiao, et~al., Study of absorption and re-emission processes in a ternary
  liquid scintillation system, Chinese Physics C - CHIN PHYS C 34 (2010)
  1724--1728.

\bibitem{LYAR}
C.~Aberle, C.~Buck, F.~X. Hartmann, S.~Schönert, Light yield and energy
  transfer in a new gd-loaded liquid scintillator, Chemical Physics Letters
  516~(4-6) (2011) 257--262.

\bibitem{TAR}
U.~T. Marrodán, et~al., Fluorescence decay-time constants in organic liquid
  scintillators, Review of Scientific Instruments 80~(4) (2009) 091302.

\bibitem{SNO}
B.~Aharmim, et~al., Measurement of the cosmic ray and neutrino-induced muon
  flux at the sudbury neutrino observatory, Physical Review D 80 (2009) 012001.

\bibitem{Atten}
P.~Huang, et~al., Study of attenuation length of linear alkyl benzene as ls
  solvent, JINST 5 (2010) P08007.

\bibitem{Atten2}
G.-Y. Yu, et~al., Some new progress on the light absorption properties of
  linear alkyl benzene solvent, Chinese physics C 40~(1) (2016) 016002.

\bibitem{Zhouxiang1}
X.~Zhou, et~al., Rayleigh scattering of linear alkylbenzene in large liquid
  scintillator detectors, Rev. Sci. Instrum. 86 (2015) 073310.

\bibitem{Zhouxiang2}
X.~Zhou, et~al., Spectroscopic study of light scattering in linear alkylbenzene
  for liquid scintillator neutrino detectors, Eur. Phys. J. C75 (2015) 545.

\bibitem{SNOplus}
M.~C. Chen, The sno liquid scintillator project, Nuclear Physics B (Proceedings
  Supplements) 145~(145) (2005) 65--68.

\bibitem{zhiyu}
Y.~Zhi, Y.~Liang, Z.~Wang, S.~Chen, Wide field-of-view and high-efficiency
  light concentrator, arXiv preprint arXiv:1703.07527.

\bibitem{15m}
L.~Gao, et~al., Attenuation length measurements of a liquid scintillator with
  labview and reliability evaluation of the device, Chinese Physics C 37 (2013)
  076001.

\bibitem{1ton}
Z.~Wang, et~al., Design and analysis of a 1-ton prototype of the jinping
  neutrino experiment, Nuclear Instruments and Methods in Physics Research
  Section A: Accelerators, Spectrometers, Detectors and Associated Equipment
  855 (2017) 81 -- 87.

\end{thebibliography}

\end{document}